# Real-Time AI-Driven Milling Digital Twin Towards Extreme Low-Latency

Wenyi Liu[a,b], R. Sharma[a,b], W. "Grace" Guo[b,c], J. Yi[a,b], Y.B. Guo[a,b*]

[a] *Department of Mechanical and Aerospace Engineering, Rutgers University-New Brunswick, Piscataway, NJ 08854, USA*
[b] *New Jersey Advanced Manufacturing Institute, Rutgers University-New Brunswick, Piscataway, NJ 08854, USA*
[c] *Dept. of Industrial and Systems Engineering, Rutgers University-New Brunswick, Piscataway, NJ 08854, USA*

*Corresponding author: yuebin.guo@rutgers.edu

## Abstract

Digital twin (DT) enables smart manufacturing by leveraging real-time data, AI models, and intelligent control systems. This paper presents a state-of-the-art analysis on the emerging field of DTs in the context of milling. The critical aspects of DT are explored through the lens of virtual models of physical milling, data flow from physical milling to virtual model, and feedback from virtual model to physical milling. Live data streaming protocols and virtual modeling methods are highlighted. A case study showcases the transformative capability of a real-time machine learning-driven live DT of tool-work contact in a milling process. Future research directions are outlined to achieve the goals of Industry 4.0 and beyond.

## 1. Introduction to Digital Twins

Smart manufacturing or Industry 4.0 represents an emerging manufacturing paradigm for improving production efficiency, adaptability, customization, and sustainability by leveraging digital technology and intelligent systems[1]. The vital enabling technologies include networked sensors, Internet of Things (IoT), Artificial Intelligence (AI) and Machine Learning (ML) (collectively termed AI), real-time data analytics[2,3], dynamic modeling and simulation, and intelligent automation[4]. The integrated sensing and communication network is connected to the manufacturing system that interacts with machines (including robots), transmits information, and makes intelligent decisions based on system algorithms[5]. By integrating the sensing and communication network, big data analytics, and AI, smart manufacturing can enable real-time monitoring, process automation, and predictive maintenance. Data is critical in smart manufacturing, transforming information from various sources into actionable decisions through comprehensive collection, storage, processing, visualization, and transmission, thereby improving production efficiency and responsiveness[6,7].

This paper aims to focus on milling, a representative manufacturing process, to explore approaches towards smart manufacturing. Milling is widely used in the automotive and aerospace industries and provides complex process dynamics. For example, progressive tool wear can lead to significant cutting dynamics such as chattering, poor surface finish, dimensional accuracy, rising energy usage, and even machine damage. It is reported that cutting tool wear accounts for nearly 75% of total production downtime, while maintenance of the machine tool's main components contributes to 12% of the overall production cost[8,9]. Implementing manufacturing process monitoring and predictive analytics is essential for assessing the condition of cutting tools, forecasting wear progression[10], reducing the need for post-process quality inspections[11].

Among the vital enabling technologies, digital twin (DT) provides an approach to achieve real-time sensing, learning, and control for smart manufacturing. A digital twin (DT, Figure 1) is a digital replica of a physical entity (e.g., process, machine, or system) with a live bidirectional connection between them[12]. The key idea of DTs lies in bridging the critical gap between a static model and a live data stream representing the dynamic physical entity[12–14]. Since the original concept[15] and terminology of DT[16,17], various definitions, understandings, and applications of DTs have evolved in vertical domains[18–26]. DTs act as dynamic, data-driven models that enable real-time monitoring, simulation, and optimization using data from various sources (e.g., IoT sensors).

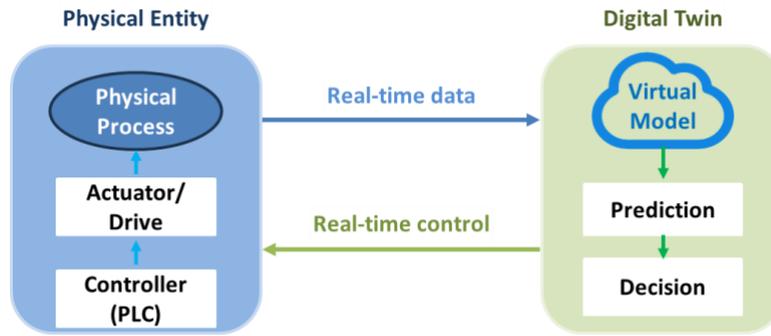

Fig. 1: Concept of digital twin (DT).

A digital replica can be represented in different types, such as physics-based simulations and data-driven models. Physics-based simulation models, including finite element analysis (FEA) and computational fluid dynamics (CFD)[27,28], are based on physical principles and mathematical equations. These simulation models are accurate and valuable for understanding the underlying physics of a process, but can be computationally expensive and slow to adapt to real-time changes. On the other hand, data-driven models, including AI/ML models, rely on data-driven approaches that learn patterns from large datasets collected through sensors and monitoring systems. These models can predict future outcomes, detect anomalies, and optimize process parameters. They are flexible and adaptable because they can be continuously improved over time through the input of real-time data, thereby providing real-time inspections of system behavior. The key feature of a DT is its ability to update and calibrate the virtual model in real time. As data is collected from various monitoring technologies, such as sensors on the milling machine, it is fed back into the DT to update the model continuously. This real-time update ensures that the virtual model accurately reflects the physical system, which is critical to maintaining the precision of the manufacturing process. For instance, if machine conditions change (such as tool wear or temperature fluctuations), these changes can be applied to calibrate the DT model, ensuring accurate prediction and control of the process. Finally, model-based process control leverages the DT to optimize and control the physical process. In this approach, the digital model's predictive capabilities, which are synchronized with real-time data, inform control decisions. For example, DTs can predict how to adjust cutting parameters, such as spindle speed or feed rate, that can affect the quality of the final product in a milling operation. The model can then recommend adjusting to maintain a stable process, ensuring high efficiency, reduced material waste, and consistent surface quality. By integrating real-time updates, predictive analytics, and automated control, DTs enable continuous optimization of the milling process and quick responses to changes in real-time conditions.

Despite the transformative capabilities DTs may generate, several key issues remain to block the development of a functional DT: (1) Virtual Representation of Physical System - Evaluation of different types of DT models in terms of accuracy, cost, efficiency, and interpretability; (2) Flow

From Physical System to Virtual Model - Synchronization a DT model with live data from the viewpoint of a physical milling to the virtual model; and (3) Feedback From Virtual Model To Physical System - Decision-Making and Real-time Process Control.

The objectives of this work are multifold. The first objective of this paper is to examine the three aspects of DT development based on a comprehensive and in-depth investigation of the state-of-the-art. The second objective is to showcase a functional milling DT to demonstrate its transformative capabilities based on the real-time acoustic emission monitoring, a neural network (NN) model, model synchronization with live data stream, and decision-making for real-time control. The third objective is to identify current challenges and provide future research directions.

## 2. Virtual Models of Physical Milling

A model is the core asset for a DT. The transition from physical milling to its virtual representation relies on various modeling techniques that simulate and predict milling operations. These models form the basis of DT systems that digitally replicate the physical milling process to facilitate real-time monitoring, predictive decision-making, and process control. Key modeling methods include physics-based simulation models, data-driven ML models, and hybrid models using both of these methods (Figure 2)[29,30]. Physics-based models rely on established principles of mechanics, material science, and thermodynamics to simulate the physical behaviors of the milling process[31]. These models provide insights into the underlying physics of the process, but they can be very computation-intensive and time-consuming. Traditional physics-based modeling methods are challenging to simulate complex nonlinear relationships, which often exist widely in real-world applications. In contrast, data-driven ML models use extensive data sets from real-world milling operations to learn patterns and predict outcomes. These models make it possible to capture nonlinear relationships. They also enable real-time updates as new data is collected, making them highly adaptable to changing conditions.

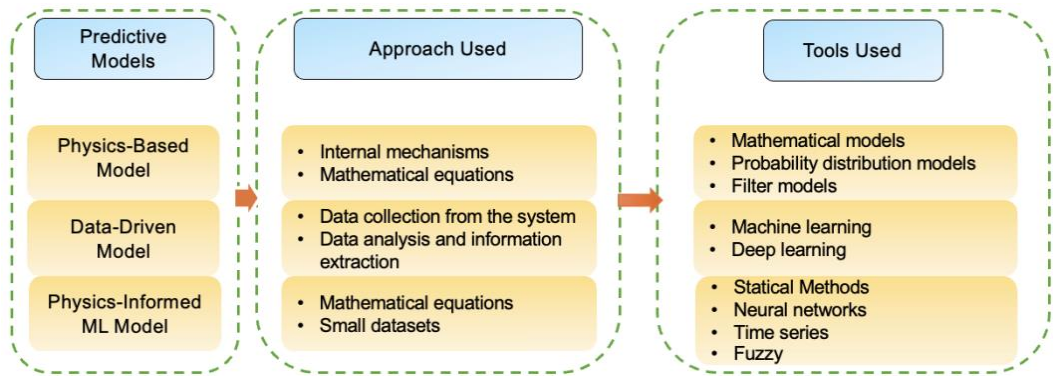

Fig. 2: Predictive models for machining tool maintenance[35]

AI/ML models are expected to play a significant role in developing future DT systems. By continuously adapting to real-time data, AI/ML models can enhance their predictions and respond to dynamic changes in the manufacturing process. This ability improves the capability of DTs to simulate and optimize milling operations in real time, leading to better decision-making, more efficient production, and reduced downtime. Integrating AI/ML into DT models promises to transform how milling processes are controlled and optimized in the future, offering the potential for autonomous decision-making, self-correction, and continuous process improvement[32–35].

## 2.1 Physics-based simulation models

Physics-based approaches are widely applied in manufacturing process optimization to meet the demands arising from the increasing complexity of parts, quality requirements, and the growing need for machining components made from materials with high machining resistance. Physics-based simulation models are significant for understanding and optimizing milling processes. Physics-based simulation models use numerical methods to simulate real-world physical processes. Physics-based models use complicated equations to get parameters like cutting forces, power consumption, torque, and material removal rate. Besides, they are also used to understand the manufacturing process and parameters that measurement cannot acquire directly.

Numerical methods are computational approaches that simulate, analyze, and optimize milling processes by solving mathematical models. These methods help predict machinability[36]. Some key numerical methods used in milling include FEA, finite difference method (FDM), and mechanistic models. These methods are often used in conjunction with each other to provide a comprehensive understanding of the milling process. FEA is the most commonly used numerical method for modeling milling, especially in industry settings where precise and optimized machining parameters are essential[37]. These models use fundamental physical principles, such as mechanics, thermodynamics, and material behavior, to simulate the interaction between tools, workpieces, and machining conditions. For instance, 3D cutting simulations based on the finite element method (FEM) play a crucial role in analyzing key physical state variables, enabling the reduction of cutting forces, friction, and tool wear in micro-textured cutting tools[38].

The advantage of physics-based milling simulation models is that they provide interpretable results by directly simulating the physical interactions between milling components (the cutter and the workpiece) and physical parameters, enabling more accurate predictions and a deeper understanding of the machining process. While the advantages of physics-based simulation models make them highly valuable in milling process studies, their limitations must also be carefully considered. The disadvantages of physics-based simulation models include the challenge of model updating to reflect real-time process changes and the high computational cost of solving complex physical equations, which can limit their practical application in some scenarios.

Variables such as load, force, and torque can be modeled mathematically to simulate and optimize the milling process. Physics-based models are also utilized to predict and mitigate chatter during milling operations by integrating dynamic cutting force coefficients and cutter geometry into time-domain simulations. Process stability and optimization strategies for spindle speeds and cutting depths for high-precision milling offer a better understanding of the strengths and limitations of current predictive performance models[29]. These models aim to establish a relationship between predicted fundamental process parameters and various performance metrics, including product quality (such as accuracy, dimensional tolerances, and surface finish), surface and subsurface integrity, chip formation and breakability, and process stability[39], and tool wear[40].

A physics-based model can also be used to predict and control chatter during milling by combining stability analysis with dynamic cutting force coefficients to model the interactions between the tool, workpiece, and machine system for real-time prediction of chatter, offering practical solutions for adjusting cutting parameters to avoid instability[41]. By considering tool deformation and thermal stress due to temperature and combining multi-physical simulation methods of thermal, mechanical, and dynamic models to optimize high-speed milling processes, a more comprehensive simulation environment has been developed to predict tool life and material behavior under different operating conditions. This integrated modeling approach offers new insights into optimizing milling for precision machining simulation[42]. Physics-based simulations are integrated with optimization algorithms to refine milling process parameters. Cutting conditions are determined by simulating the milling process and evaluating the effects of variables like cutting force, surface finish, and tool wear. The physics-based approach optimizes machining processes and enhances production rates while maintaining product quality[43]. Table 1 compares various physics-based simulation methods for milling, detailing their approaches to cutting forces, temperature distribution, stress-strain analysis, and tool wear, with hybrid models integrating multiple techniques for improved accuracy.

Table 1: Comparison of different physics-based simulation methods.

| Milling Parameter | Finite Element Method (FEM) | Finite Difference Method (FDM) | Smoothed Particle Hydrodynamics (SPH) | Hybrid Models |
|---|---|---|---|---|
| Cutting Forces | Calculated using stress-strain analysis and force equilibrium equations | Applicable in simplified Models | Calculated through particle interactions and momentum conservation | Combines FEM and SPH for accurate force prediction |
| Temperature Distribution | Determined through heat transfer modeling and thermal-mechanical coupling | Computed using numerical heat conduction equations | Determined by tracking thermal diffusion among particles | Uses FEM for bulk material heating and SPH for localized effects |
| Stress and Strain | Derived from deformation analysis using constitutive material models | Estimated based on numerical differentiation of strain fields | Derived from particle-based stress-strain relationships | Integrates multiple approaches to improve strain accuracy |
| Tool Wear | Predicted using contact mechanics and tool-workpiece interaction models | Requires additional wear models | Modeled dynamically by tracking tool-workpiece interactions | Enhances prediction accuracy by incorporating multiple wear mechanisms |

## 2.2 Data-driven ML models

A data-driven model is a mathematical, statistical, or ML model built and trained using empirical data to identify patterns, make predictions, or make decisions directly from the data, without modeling the system based on physical laws. To ensure data reliability, data curation processes such as normalization and down-sampling are often applied to improve the input data quality and reduce measurement noise. Data-driven models have been used to predict tool wear, chatter, cutting force, and surface topography. To improve prediction accuracy under different cutting conditions, these models often combine multimodal sensor data with process parameters such as spindle speed, feed rate, and cutting depth. In the context of milling operations, these variables can be collected through sensors and metrology systems, transformed into representative feature sets, and analyzed using ML models[44]. A wide range of ML techniques, including statistical models, NNs, time series analysis, and fuzzy logic, have been employed in smart milling to predict tool wear, chatter, and surface roughness[45]. Standard ML methods, such as Support Vector Machines (SVMs), Decision Trees, K-Nearest Neighbors (KNN), and Convolutional Neural Networks (CNNs), are used to predict process performance. Studies have shown that ML can successfully predict tool wear in milling processes[46]. SVM is a supervised ML algorithm for classification and regression tasks. It operates by identifying the optimal hyperplane that separates data points into distinct classes, making it especially effective for binary classification scenarios[47]. While Decision Trees provide interpretable decision rules and effectively identify key parameters influencing surface roughness[48], KNN offers simplicity and competitive performance when classifying spindle speed-axial depth combinations into stable or unstable conditions[49]. Neural network (NN) is the most classical ML form, providing the conceptual basis for modern deep learning architectures[50]. Those models offer valuable insights into tool wear, cutting forces, and surface quality based on large datasets. These methods are particularly useful for real-time monitoring, predictive maintenance, and improving machining efficiency by analyzing sensor data such as vibrations, sound, and cutting force[51–53].

The type and characteristics of the data sets play a critical role in guiding the choice of models[54]. For example, CNNs are designed for processing spatial data, such as images or topographic surface maps, and have been applied in smart milling for surface defect detection and image-based tool wear evaluation. In contrast, Recurrent Neural Networks (RNNs) are specifically designed for processing sequential data and perform well in modeling temporal dependencies in processing signals, thereby predicting tool wear evolution and chatter behavior more accurately. Since force, vibration, acoustic emission, temperature signals, and tool wear images reflect different milling process characteristics, different ML models should be adopted accordingly. Those data capture distinct aspects of the milling process, requiring different modeling strategies. For instance, SVM can be applied to tool wear condition monitoring[55], chatter detection[56], and surface roughness prediction[57], while a NN tends to perform better in predicting cutting forces, as it can model nonlinear relationships among multiple process variables[29,58]. Decision Trees helped identify key parameters affecting surface roughness[48], and KNN can be used to classify stable and unstable spindle speed-axial depth combinations[49]. Traditional statistical models focus on inference and understanding relationships, while ML models focus on prediction and optimizing performance. Statistical models provide better interpretability than ML models, while ML models handle nonlinear and high-dimensional data more effectively. Data characteristics, such as dimensionality, sampling rate, and noise level, determine an ML model's suitability and prediction accuracy, where high-dimensional, high-rate data favor deep networks, while traditional statistical models better handle low-dimensional or noisy data. In addition, poor-quality or imbalanced data may lead to overfitting and reduced generalization, thereby affecting an ML model's predictive reliability. Therefore, model selection should balance accuracy, interpretability, and computational efficiency based on the different demands of milling applications. Features extracted from time, frequency, and time-frequency domains, such as peak value, Root Mean Square (RMS), and energy, characterize the underlying signal behavior and determine how effectively an ML model can learn and process patterns. The backpropagation algorithm optimizes an ML model by minimizing a loss function between the actual and desired outputs, such as Mean Squared Error (MSE), Mean Absolute Error (MAE), Root Mean Square Error (RMSE), or Mean Absolute Percentage Error (MAPE)[59]. These data-driven methods may complement each other to achieve process monitoring and optimization. They can reveal the complex relationship between the extracted feature set and the corresponding parameters, like tool wear level, thereby achieving robust and accurate prediction during machining processes[60–64].

The recent development of AI has led to architectures with reasoning and multimodal understanding capabilities, including large language models (LLMs), Transformers, and reinforcement learning (RL). LLMs enable the semantic interpretation of manufacturing data for adaptive decision-making[65]. A Transformer is an NN architecture that utilizes self-attention mechanisms to process input data in parallel, especially sequential data like text, rather than step-by-step as in previous models like RNNs or Long Short-Term Memory (LSTM) networks. This architecture improves the model's generalization performance and real-time prediction ability under dynamic processing conditions[66–68]. Meanwhile, RL is a powerful approach for adaptive process control, allowing its agent to autonomously learn optimal machining strategies through continuous interaction with the physical system[69]. These AI technologies have jointly driven the development of cognitive and adaptive digital twin systems, laying the foundation for achieving intelligent, low-latency, and autonomous milling operations.

Overall, data-driven models demonstrate excellent adaptive capabilities in complex processing dynamics modeling and can support online learning and predictive maintenance through real-time feedback. These models perform exceptionally well under sufficient training datasets and stable operating conditions. However, the prediction accuracy often declines when applied to unknown or highly dynamic milling conditions if material properties, tool geometry, or environments significantly differ from the training domain. Currently, most methods rely on a large amount of labeled data, increasing the computational cost. Therefore, the application scope of these models is usually limited to processing conditions with clear features or repetitive patterns. Future research could focus on developing emerging ML technology, like Transformers and RL, combined with the milling process physics to achieve intelligent modeling with interpretability, portability, and computational efficiency for intelligent digital twins of milling systems.

## 2.3 Physics-informed ML models

Physics-based ML (PIML) methods, also called hybrid modeling methods, combine physics-based models with data-driven techniques to improve prediction accuracy while ensuring consistency with the laws of physics. This method takes advantage of both methods to achieve reliable and robust results. Instead of relying solely on data, PIML enforces physical laws (such as conservation of mass, momentum, and energy) as constraints within neural networks[70]. Data-driven ML models perform well but often act as "black boxes," making them hard to interpret[71]. Combining physics-based and data-driven approaches, hybrid models are built to enhance the accuracy of milling simulations. Physics-based models ensure physical consistency, while data-driven elements improve the ability to adjust, modify, or change in response to new conditions. This approach bridges the gap between traditional modeling techniques and modern ML[29]. PIML models have three primary purposes for smart machining: 1) Hybrid models combine data-driven and physics-based approaches to enhance model performance; 2) Physics-guided loss functions integrate prior knowledge or physics models into regularized terms, ensuring physical consistency; 3) Physics-pretrained hybrid models use constraints during model initialization, which accelerates training and improves consistency[72]. Unlike transfer learning, PIML incorporates physics directly into the learning process instead of transferring knowledge from a pre-trained model. However, PIML could be combined with transfer learning by pretraining a model on one material and adapting it to another while maintaining physical consistency[73]. The hybrid approach reduces dependency on large datasets and ensures that predictions align with real-world physics methodology.

PIML models in milling have significant advantages due to combining a data-driven approach with established physical principles. These models can make accurate predictions with relatively small data sets because they utilize physical and mathematical models, reducing the need for large amounts of experimental data. Additionally, they enhance the evaluation of model uncertainty, ensuring more reliable and robust results. However, these models also face some challenges. Model integration can be technically complex, requiring domain knowledge in both machining physics and ML. Furthermore, computational costs can be high, particularly when solving embedded physical equations or dealing with complex multi-physics interactions. These limitations may reduce the scalability and adaptability of PIML models in diverse or large-scale industrial applications.

## 3. Flow From Physical Milling to Virtual Model: DT Model Updating with Live Data

Data acquisition and live data stream are essential inputs for a milling process DT. Data acquisition via process monitoring in milling involves various techniques to track milling process performance and tool condition. Traditional monitoring techniques, including acoustic emission, vibration, force, and temperature measurements, each provide critical data for understanding the milling process. Live data stream for DT model updating and calibration involves multimodal data integration, where model data interfaces facilitate data input, model update, output, and feedback loops to improve predictions and increase accuracy. Besides, 5G-based process monitoring enables high-speed and ultra-low latency data transmission, enabling real-time data collection and integration. The following sections will discuss the role of data acquisition and live streams in process monitoring and DT deployment, focusing on sensor-based data acquisition, real-time data processing, online predictions, model updating, and the impact of 5G technology in smart milling.

### 3.1 Data acquisition and visualization

Different types of sensors are used to collect data during milling, including acoustic emission (AE), force, vibration, and temperature sensors, which will be introduced in the section below. The integration of data from these diverse sources forms multimodal data, which provides a comprehensive representation of the milling process. Table 2 highlights the diverse sensor types used in milling. Integrating multimodal data from these sensors enables a more thorough understanding of tool conditions, material behavior, and milling dynamics, enhancing predictive modeling and real-time decision-making in smart milling.

Table 2: Sensors and features.

| Sensor Type | Measured Parameter | Equipment | Advantages | Challenges |
|---|---|---|---|---|
| Acoustic Emission (AE) | Elastic waves | Preamplifiers, amplifiers, filters, piezoelectric AE sensors, signal analyzers | Real-time monitoring, high sensitivity, ability to detect microcracks | Requires signal processing expertise, sensitive to ambient noise |
| Force | Cutting forces | Piezoelectric sensors, rotary dynamometers, triaxial force sensors | Reliable tool condition monitoring, strong correlation with surface roughness | Expensive commercial dynamometers, high data acquisition cost |
| Vibration | Tool and workpiece oscillations | Accelerometers, vibration analyzers, computer-based real-time signal processing | Low cost, can detect chatter and tool imbalance effectively | Difficult to filter signals, affected by sensor placement and cutting fluid |
| Temperature | Temperatures | Thermocouples, infrared sensors, IR thermal imagers, microcontroller-based monitoring systems | Non-contact measurement, fast response, suitable for high-speed machining | Affected by environmental factors, requires calibration for accuracy |

Sensors output the generated signals, and data acquisition (DAQ) systems are widely used in data collection. The first stage of data collection is gathering raw data from sources such as sensors. The original data is the primary input, so it is essential to collect data adequately to ensure its accuracy, completeness, and relevance. The next step is data preparation, which aims to generate accurate, comprehensive, and relevant data sets to support subsequent processing steps. After connecting the specific sensors, which are applied to gain needed physical parameters, to the preamplifier and DAQ and configuring the sampling rate and data format, data will then be delivered to the central processing unit without any loss[74]. A preamplifier is an electronic device designed to amplify weak electrical signals into stronger ones, filter noisy signals to clearer output signals, and support further processing. Otherwise, the resulting signal would likely face the problem of noise or distortion. Preamplifiers are commonly used to enhance signals from analog sensors, and to minimize the impact of noise and interference. They are typically positioned near the sensor.

A DAQ system comprising the required hardware and software is designed and integrated to enable the automated reading and storage of sensor data on a computer during metal cutting[75]. The Analog-to-Digital Converter (ADC) is a fundamental component of modern DAQ systems, responsible for converting analog signals into digital data that can be transmitted, stored, and analyzed as required. The potential difference generated by the sensor is fed into the embedded data acquisition unit via an amplification circuit. The embedded unit then converts the signal into a digital format and transmits it to the computer, where it is subsequently displayed and further processed[76]. The DAQ software can perform real-time graphical simulations of signals during the process[77], providing immediate visualization and analysis. Alternatively, this functionality can be achieved using Application Programming Interfaces (APIs), enabling seamless integration with other data processing systems.

## 3.2 DT synchronization and calibration

A live data stream for DT synchronization and calibration is critical to maintaining accurate and adaptive machining models. An efficient data pipeline is essential for seamlessly collecting, processing, and transforming sensor data for a milling DT. A data pipeline consists of four key stages: ingestion, storage, transformation, and aggregation[78]. Data streaming is a subset of a data pipeline and focuses specifically on real-time data flow[79]. A real-time data pipeline often integrates a data streaming system (e.g., Kafka, Flink, Spark Streaming) to handle high-speed data from sources such as IoT sensors, logs, or financial transactions. Data streaming enables continuous data transmission for real-time monitoring, model updating, prediction, and decision-making. In addition, by being updated and calibrated based on online data, the model can maintain accuracy and adaptability. This section explores the end-to-end data streaming process and iterative updates to enhance DT performance and process optimization.

### 3.2.1 Live data stream

After data is collected and preprocessed, it can be transmitted to the cloud platforms or edge servers for further data processing. This data needs to be efficiently transmitted for additional processing, analysis, and storage. Data streaming continuously transmits data from various sources in real time. In smart milling and industrial IoT, real-time streaming data is crucial for timely predictions, decision-making, process optimization, and ensuring precision in operations by continuously monitoring tool wear, vibration, and cutting forces[80]. Different communication protocols can be used for real-time data transmission depending on latency, reliability, scalability, and interoperability requirements to achieve real-time data transmission. For example, Message Queuing Telemetry Transport (MQTT) protocols facilitate seamless communication between networks and devices and enable low-latency, real-time communication[81,82]. OPC UA (Open

Platform Communications Unified Architecture) is another standard protocol used in industrial automation for machine-to-machine communication[83]. It provides a secure, platform-independent standard for integrating multi-vendor industrial devices with ML models[84]. Besides, Constrained Application Protocol (CoAP) uses the REST philosophy and is proposed for device communication, especially when many sensors and devices are within the network[85,86]. In addition, WebSocket provides a complete bidirectional communication channel through a single socket, allowing both sides to send data at any time while the connection is established[87].

In addition to communication protocols, various data platforms such as Apache Kafka and Redpanda can be deployed on servers to enable data streaming and efficiently handle high-throughput sensor data in smart milling applications. Take Apache Kafka as an example. It is widely used in industrial applications because it can ingest, process, store, and distribute real-time sensor data with low latency. Apache Kafka is a widely used open-source distributed streaming platform commonly adopted for building large-scale data streaming applications, such as those used by LinkedIn[88,89]. Kafka uses an API (Application Programming Interface) and its own binary protocol over Transmission Control Protocol (TCP) for communication. Its architecture follows a publish-subscribe model, where producers send data streams to Kafka topics, brokers store and manage data, and consumers retrieve and process the data as needed[89,90]. This structure allows for high-throughput parallel processing, ensuring that real-time manufacturing data can be analyzed efficiently. For instance, acoustic emission and force sensors continuously act as Kafka producers in a milling process, sending real-time machining data to Kafka topics, which are then stored and distributed across multiple brokers. AI-driven monitoring systems and DT models subscribe to these topics and consume data for anomaly detection, tool condition monitoring, and process optimization. Kafka ensures fault tolerance through data replication across multiple brokers, preventing data loss and maintaining system reliability even if a node fails. Data streaming platforms use communication protocols to collect real-time production data from machines, which can be integrated with APIs to enable seamless data transmission between applications, integrating with cloud computing platforms like AWS, Azure, or Google Cloud for remote monitoring and predictive maintenance. Table 3 compares different communication protocols and data streaming platforms for DT applications.

Table 3: Data streaming protocols[83–104]

| Name | Type | Communication Mechanism | Advantages | Disadvantages |
|---|---|---|---|---|
| MQTT | Lightweight messaging protocol over TCP/IP | Brokered publish/subscribe using a lightweight header format | Lightweight and low overhead; reliable over unreliable networks; ideal for constrained IoT devices | Limited advanced stream processing capabilities; not built for high-volume data |
| OPC UA | Industrial communication protocol over TCP/IP | Client-server and publish/subscribe with integrated semantic modeling and security | Rich information modeling; standardized security and reliability; suitable for industrial device communication | High complexity and overhead; requiring more computational resources, not optimal for low-resource devices |
| CoAP | RESTful web transfer protocol over UDP | Request-response model, follows a web-like architecture | Lightweight; fast in constrained networks; suitable for constrained sensor networks | Limited reliability due to use of UDP; limited scalability, not widely used by industrial or enterprise systems |
| WebSocket | Full-duplex communication protocol over TCP | Real-time browser/server, bi-directional stream | Real-time bidirectional communication; appropriate for web and mobile real-time apps | No standard message format; manual reconnection and state handling required |
| Apache Kafka | Distributed streaming platform over TCP | Log-based publish/subscribe | Durability, scalability, and message ordering; strong integration with stream processing frameworks | Heavy resource usage; complex setup and operational overhead; not ideal for low-power devices |
| RedPanda | Kafka-compatible streaming platform over TCP | Kafka API-compatible platform, log-based publish/subscribe | Lower latency due to optimized architecture, easier operations, and lower resource use | Smaller community and ecosystem; enterprise support still growing; less open-source visibility |

In real-world industrial applications, selecting the proper protocol or architecture depends on the nature of the data, communication requirements, and system constraints. MQTT is ideal for lightweight, real-time telemetry in constrained networks or edge devices, such as vibration sensors on rotating equipment. MQTT can enable seamless data exchange between three-axis CNC machines and centralized monitoring systems in a smart factory setting, allowing real-time updates of machine status and operations. This lightweight protocol bridged different machine types and eliminated the need for complex protocol conversions. Besides, edge devices like Raspberry Pi collect and preprocess sensor data locally, then publish results via MQTT to cloud services and dashboards. This architecture reduced network load and latency while enabling real-time visibility and control on the shop floor[105]. OPC UA is better suited for structured, hierarchical machine data in environments requiring tight integration with industrial automation systems, such as retrieving Programmable Logic Controller (PLC) data from CNC machines or Supervisory Control and Data Acquisition (SCADA) systems. For instance, in a laboratory-scale smart manufacturing system, OPC UA can enable seamless real-time data exchange between robotic arms and conveyor systems by acting as the communication layer. It allowed diverse devices using different physical networks (Wi-Fi) to share data through a unified OPC UA information model. In addition, A Java-based OPC UA client interface in the enterprise layer allows operators to remotely monitor system states and issue control commands via a Graphical User Interface (GUI). The OPC UA server aggregates data from field devices. It exposes it as structured information models, supporting read/write operations that facilitate real-time visualization, operational decisions, and dynamic reconfiguration of the manufacturing process[106]. Kafka is preferred for handling high-throughput, scalable data pipelines, especially when the goal is to support multiple downstream consumers such as analytics platforms, dashboards, or ML inference engines. For example, Apache Kafka enables real-time data collection from smart factories across Korea, China, and cloud platforms by linking isolated factory networks into a unified streaming pipeline. It can handle diverse sources such as sensor logs and PLC outputs, supporting global-scale integration[107].

Several communication protocols are commonly used in smart manufacturing, including MQTT, CoAP, OPC UA, and WebSocket, each directly impacting system latency. Lightweight protocols such as MQTT and CoAP generally provide low latency and minimal overhead, offering good choices for edge communication in smart milling. MQTT is reliable and efficient for constrained networks, while CoAP is faster but less robust because it uses User Datagram Protocol (UDP). In contrast, OPC UA offers strong semantic modeling and industrial integration, bringing higher latency and computational overhead. Therefore, it is better suited for structured data exchange rather than high-frequency streaming. WebSocket supports low-latency, two-way communication, which is helpful for dashboards and operator interfaces, but lacks standard industrial messaging.

Combining communication protocols and data streaming platforms can be an effective strategy for implementing a DT in the milling process. For instance, Apache Kafka can be deployed in industrial environments to enable scalable, real-time data streaming for cloud-based applications. At the same time, MQTT, as a lightweight messaging protocol, ensures reliable communication in bandwidth-constrained and unreliable network conditions. By using MQTT to interface with and collect data from distributed edge devices or sensors and leveraging Kafka to transport and process this data in real time, the system benefits from efficient device-level communication and robust, scalable back-end analytics. By integrating MQTT for reliable data acquisition at the device level with Kafka for centralized stream processing and analytics, this architecture enables timely decision-making, predictive modeling, and closed-loop control, which are critical for realizing a responsive and accurate DT in smart manufacturing systems.

Following data streaming, the next step is stream processing. Stream processing involves applying algorithms or rules to the data stream as it flows through the system, allowing for immediate analysis, aggregation or compression, filtering, feature extraction, and selection, which will be finished on servers through cloud/edge computing or streaming engines. Server computing enhances the data processing, storage, and computational capabilities of Wireless Sensor Networks (WSNs) by offloading intensive operations from local devices[108]. Since data streams are inherently unbounded, windowing allows data to be grouped based on time intervals or event-based triggers. Windowing divides a dataset into discrete segments, enabling processing to be performed on each segment as a group[109]. Stream processing frameworks apply windowing techniques that segment continuous data streams into finite chunks to address this, enabling more effective real-time analytics. Common windowing techniques include fixed windows, which divide data into equal time segments; sliding windows, which allow overlapping segments for capturing trends; and session windows, which group data based on event activity rather than time (Figure 3)[110].

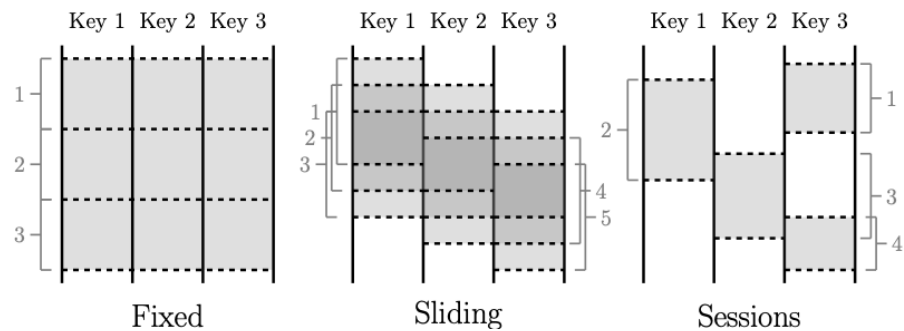

Fig. 3: Common window patterns[110]

Feature extraction and selection from sensor data in machining involves several key steps. First, relevant parameters are identified based on the specifics of the manufacturing process. Next, time-series analysis techniques are applied to extract distinct types of features, including time-domain features (mean, standard deviation, peak values), frequency-domain features (Fourier Transform for dominant frequency analysis), and time-frequency domain features (Short Time Fourier Transform or Wavelet Transform for transient or unstable signals detection). Feature selection methods such as correlation analysis, statistical tests, or ML-based techniques (e.g., Principal Component Analysis (PCA), mutual information analysis) are then used to retain the most significant features for predictive modeling. The process must consider factors such as the sensor type, the targeted outcome (e.g., defect detection or wear prediction), and the characteristics of the manufacturing process. The time, frequency, and time-frequency domains provide different perspectives for signal analysis[35]. The time domain reveals how a signal changes over time, providing direct physical interpretations. The frequency domain highlights dominant frequency components using techniques like the Fourier Transform, making it useful for detecting periodic patterns in machining processes. The time-frequency domain allows the simultaneous analysis of time and frequency characteristics using STFT or Wavelet Transform, making it particularly effective for identifying sudden tool failure or unstable process conditions. Each domain provides a unique perspective, and their combined use enhances process monitoring and predictive accuracy.

Once the most relevant features are selected, they are used to train and optimize the model. For example, Table 4 shows 12 standard AE features related to tool conditions, which are extracted in real time for further predictions. These extracted features, represented as time-series signals containing a combination of time, frequency, and time-frequency domain information, serve as inputs for ML models, enabling more precise tool condition monitoring and failure prediction[111,112].

Table 4: Definition of AE features[111]

| Features | Abbreviation | Definition |
|---|---|---|
| Rise time | RT | The time between an AE hit starts, and it reaches the peak amplitude. |
| Counts | C | The number of AE signal excursions over the AE threshold. |
| Amplitude | A | $A = 120 \log V_{max} - P$ (dB), where $P$ is preamplification gaining. |
| Root mean square | RMS | $RMS = \sqrt{\frac{1}{N}\sum_{i=1}^{N} V_i^2}$. |
| Average signal level | ASL | $ASL = 120 \log \bar{V}$ (dB) |
| Counts to peak | CP | The number of C between its start and peak amplitude. |
| Signal strength | SS | $SS = \frac{1}{f_s}\sum_{i=1}^{N}(V_i + V_{i+1})$, where $f_s$ is sample rate. |
| Absolute energy | ABE | $ABE = \frac{1}{10k\Omega}\sum_{i=1}^{N} V_i^2$, where $10k\Omega$ is the reference resistance of the recording equipment. |
| Average Frequency | AF | $AF = C/HT$, where HT is the duration of an AE hit. |
| Reverberation Frequency | RF | $RF = \frac{C-CP}{HT-RT}$. |
| Initiation frequency | IF | $IF = CP/RT$. |
| Frequency centroid | FC | $FC = \frac{\sum f \cdot \bar{V}}{\sum \bar{V}}$ is calculated from fast Fourier transform (FFT), where $\bar{V}$ is the magnitude of FFT element and f is corresponding frequency. |

To achieve the above data processing in real time, cloud/edge computing or streaming engines can be employed to enable low-latency data ingestion, processing, and delivery across systems. Cloud/edge computing operates through a combination of software frameworks, containerized applications, and runtime environments that enable real-time data stream processing across distributed nodes. Data streaming platforms can be integrated with stream processing engines like Apache Flink, Apache Spark Streaming, Azure Stream Analytics, or Java libraries like Kafka Streams to achieve real-time data processing. Stream processing engine allows users to write real-time queries on Kafka streams, making it easier to analyze and process data without writing code. For instance, Apache Kafka can be integrated with distributed computing frameworks such as Kafka Stream and Apache Spark to enhance real-time sensor data processing further. Apache Spark provides an in-memory computing framework that supports high-speed data analytics through Spark Streaming, allowing real-time manufacturing data processing.

*3.2.2 DT updating and calibration*

As shown in Figure 1, a DT (i.e., virtual model) has a real-time bidirectional connection with a physical system. Logically, the data stream (Section 3.2.1) is now ready to synchronize the DT with a selected protocol. This section addresses DT updating and calibration with the real-time data stream, while the subsequent Section 4 focuses on DT-powered real-time control of the physical system. Therefore, the three sections are seamlessly connected, which manifests the key innovation of this work. Traditional models (e.g., finite simulation model) cannot achieve real-time process learning as these on-the-fly models cannot take live process data. Instead, an ML-driven DT model can fulfill this function, which is the key innovation and beauty of intelligent DTs.

Online prediction is a significant function of the physical-to-virtual process. After feeding the processed data into ML models, predictions and decisions will be made to achieve control of the machine tool. Deploying ML models within a DT framework in a production environment involves a comprehensive pipeline to ensure real-time adaptability, scalability, and automation. The inference pipeline processes real-time data through servers, feeding pre-processed features to deployed models. ML models in this framework can be deployed through two primary methods. One method integrates models into streaming platforms such as Apache Kafka, allowing them to ingest real-time data, perform immediate inference, and output predictions directly into streaming pipelines. This low-latency, event-driven approach is well-suited for high-throughput industrial applications, including anomaly detection, predictive maintenance, and real-time process optimization in smart manufacturing. Alternatively, models can be deployed via APIs or gRPC, enabling external systems like cloud platforms, edge devices, or web-based monitoring interfaces to send data requests and receive predictions in real time. While RESTful APIs ensure broad interoperability with web services, gRPC provides more efficient, low-latency communication, making it ideal for high-speed industrial environments requiring rapid data exchange and decision-making.

Classification algorithms are essential for classifying data into discrete categories. Support vector machines (SVMs) can find the best hyperplane for separating categories, performing well in high and small data sets. Decision trees (DT) provide interpretable tree-like structures but need pruning to avoid overfitting, while random forests (RF) use ensemble learning to improve robustness. Artificial Neural Networks (ANN) model complex relationships but are computationally intensive, and K-Nearest Neighbors (KNN) classify based on proximity, though it can be computationally expensive for large datasets.

Regression algorithms, on the other hand, predict continuous values. Auto-Regressive Models are used for time series data, while Gaussian Process Regression (GPR) offers probabilistic predictions. Multiple Linear Regression assumes linear relationships, with Ridge Regression (RR) and Lasso Regression adding regularization to prevent overfitting and enhance feature selection. Principal Component Regression (PCR) reduces dimensionality before regression.

Deep learning (DL) algorithms tackle complex data. Artificial Neural Networks (ANN) are the basis, while Convolutional Neural Networks (CNN) are dedicated to image processing, and Recurrent Neural Networks (RNN) are used to model sequential data. Advanced models such as Generative Adversarial Networks (GAN) generate realistic synthetic data, demonstrating the adaptability and power of deep learning. Together, these methods address diverse ML challenges.

To ensure the accuracy and adaptability of predictive models in smart manufacturing, model updating, and calibration are critical processes that maintain the reliability of ML models over time. As sensor data continuously evolves due to process variations and environmental factors, models must be systematically refined to mitigate concept drift, which occurs when the statistical properties of the input data change. Model updating incorporates new data to retrain or fine-tune the existing model, employing incremental learning for continuous adaptation or batch retraining at predefined intervals. In real-time applications, online learning and reinforcement learning techniques can dynamically adjust models based on machining conditions, enhancing responsiveness and predictive accuracy. Calibration ensures that model predictions align with observed values by adjusting parameters based on deviations between predicted and measured outcomes. This process may involve hyperparameter optimization through Bayesian optimization or grid search and domain adaptation techniques that enable models to generalize across different machining environments or material properties. In DT systems, calibration often integrates sensor fusion, where multiple sensor modalities are combined to enhance prediction accuracy, feedback loops, and real-time deviations between predicted and actual machining performance guide model refinement.

Figure 4 presents a real-time data streaming architecture integrating MQTT and Apache Kafka for sensor-to-ML model communication. Sensor data is transmitted via MQTT to a central server, where a Kafka producer ingests the messages into a Kafka topic. The data then undergoes processing before being consumed by a Kafka consumer and passed to ML models for inference. The proposed data pipeline leverages the lightweight communication protocol MQTT and the large-scale messaging platform Apache Kafka through APIs. This integration facilitates scalable, low-latency data streaming, making it suitable for real-time DT implementations and predictive maintenance in smart milling systems. Automated pipelines can be deployed using Apache Kafka and Spark Streaming to enable seamless model updating and calibration, ensuring continuous model adaptation without manual intervention. The data ingestion layer captures real-time sensor data. Once collected, the data processing layer transforms raw sensor signals into meaningful features essential for ML models. These extracted features are fed into the model inference layer, where trained ML models perform real-time predictions, facilitating proactive decision-making. The model updating layer employs incremental learning and batch retraining to maintain accuracy and adaptability, allowing models to evolve with new data and remain robust against process variations. Finally, the visualization and integration layer provides real-time monitoring and decision support, ensuring seamless integration with manufacturing control systems. Together, these interconnected layers establish a comprehensive, real-time data-driven framework that enhances adaptive manufacturing, ensuring predictive models remain accurate, reliable, and responsive to dynamic production conditions[113].

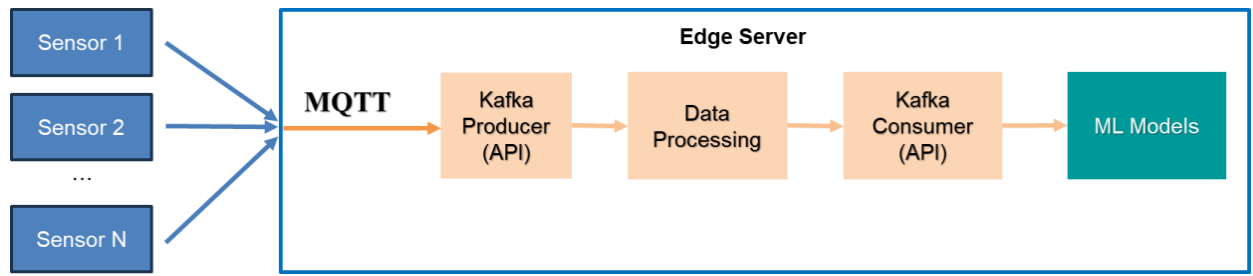

Fig. 4: An example of data acquisition and DT model deployment architecture.

Additionally, the combination of OPC UA, MQTT, and Kafka can form a robust and efficient architecture for IoT applications by aligning communication and processing capabilities across different layers of a smart milling system. MQTT ensures efficient, real-time data acquisition from distributed edge devices, particularly in bandwidth-constrained or unreliable network environments. OPC UA provides structured access to machine-level data, enabling seamless integration with industrial equipment such as CNC controllers and PLCs. Kafka acts as the central streaming platform, aggregating data from MQTT and OPC UA sources, and enabling real-time analytics, predictive modeling, and DT synchronization. This architecture allows a smart factory to collect sensor data via MQTT, retrieve machine status via OPC UA, and stream both into Kafka for unified processing. This setup supports scalable data integration across heterogeneous systems, ensures timely insights for predictive maintenance and quality control, and enables closed-loop decision-making. By assigning each protocol to the layer where it performs best, device-level, machine-level, and platform-level, the system is optimized for performance, interoperability, and responsiveness in modern IIoT environments[114–116]. For example, a smart factory might use OPC UA to interface with machine tools, MQTT to gather sensor data from production lines, and Kafka to unify and stream both types of data into a centralized analytics platform for predictive maintenance or quality monitoring. Choosing the right combination ensures each system layer is optimized for performance, reliability, and flexibility.

## 4. Feedback From Virtual Model to Physical Milling: Decision-Making & Real-Time Control

After completing real-time predictions, integrating the AI model outputs into the digital twin enables translating model predictions into real-time control actions for the physical milling operations to achieve adaptive milling. In many systems, ML model predictions remain isolated from direct actuation, preventing real-time adjustment of machining parameters. Linking prediction and execution allows the milling system to adapt continuously to changing conditions. When parameters such as feed rate and spindle speed are adjusted based on real-time predictions, the pure monitoring framework evolves into an adaptive system capable of self-optimization and low-latency operation.

### 4.1 AI model for decision-making and real-time process control

AI models predict various aspects of the milling process, such as estimating the remaining useful life of cutting tools (tool wear prediction), assessing the quality of finished parts (surface quality prediction), and monitoring cutting forces to prevent overload or deflection (force/torque prediction). Once these predictions are made, the next step is to convert them into actionable control commands that control the system's operation. ML models can transform the prediction results into real-time control algorithms to adjust the machining process precisely. With ML, the system can dynamically correct errors, reduce risks, and effectively avoid dangerous situations. Meanwhile, ML improves the accuracy of decision-making by analyzing complicated data, ensuring more reliable processing results. These technologies make the milling process more efficient and flexible, significantly improving productivity, safety, and process stability[117,118].

Once AI model predictions are generated, these predictions must be systematically translated into actionable control commands through decision-making processes. A threshold-based approach is commonly employed when the AI model provides categorical outputs. For instance, if tool wear is predicted to exceed 80% of its usable lifespan, the system can trigger tool replacement or adjust machining parameters, such as reducing feed rate or spindle speed, to mitigate further wear. Similarly, surface quality predictions, such as excessive surface roughness, can prompt adjustments to machining parameters to improve the finish. Anomalies, such as tool breakage or chatter, may necessitate halting the machining process or reducing cutting forces to protect the machine and workpiece. In cases where continuous values are provided by the AI model (e.g., wear levels or cutting force trends), real-time optimization can be performed, dynamically optimizing parameters such as feed rate and depth of cut to maintain machining stability and ensure operational efficiency[119,120].

For more advanced applications, Model Predictive Control (MPC) can be employed to integrate AI predictions into an optimization-based control framework. MPC predicts future process states using real-time inputs such as spindle speed, feed rate, and cutting depth while considering system constraints like machine torque or thermal limits. The objective is to achieve operational goals such as minimizing tool wear, maximizing productivity, and maintaining part quality. For example, if tool wear is predicted to accelerate, MPC can optimize feed rates and spindle speeds to extend tool life while maintaining production targets[121–123]. These optimized decisions are implemented via interfaces such as Computer Numerical Control (CNC) or Programmable Logic Controllers (PLC), which convert AI-driven commands into machine actions[124]. A control submodule autonomously manages the milling process using validated NC codes (such as G-codes). These codes, containing optimal milling parameters, are verified through high-fidelity simulations, assessing factors like machine tool over-travel, tool contact, machining time, spindle current, and milling force. Once validated, the codes are sent through industrial communication protocols for process control, with updates from the optimization submodule enabling real-time adjustments to minimize deformation in thin-walled parts. The monitoring submodule tracks real-time deformation during milling, primarily influenced by milling forces. A triaxial force sensor collects force data, mapping the average force per NC code line to an instruction domain. This mapping, detailed in an Extensible Markup Language (XML)-based parameter table, evaluates each NC code's impact on deformation[125,126].

These actions bridge the gap between AI predictions and practical industrial applications. The gap here refers to the method of translating an AI model prediction into real-time control action for the physical milling operations to achieve adaptive milling. Under real-time control, controllers adjust the milling process based on AI-driven commands. For instance, the sensor feeds back the spindle load generated during the machining process to the controller, which compares the reference value with this signal and outputs a new feed command to stabilize the spindle load[127]. Some load milling controllers work under software environments like LabVIEW[128]. The designed controller is programmed in LabVIEW and executed in the NI processor. The amplifier outputs a control voltage through voltage gain to drive the actuator to suppress chatter[129]. The nonlinear controller gives the actuator a nonlinear output, which helps suppress high-frequency vibration. The controller sends signals to an actuator to perform the physical movement based on those signals. The electromagnetic actuator is a common actuator employed in the milling process. The electromagnetic force is controlled by controlling the voltage and frequency to control the operation of the machine tool[130,131]. Sensors within the system monitor these adjustments and relay feedback into the control loop, ensuring that the implemented changes yield the desired outcomes. This closed-loop system creates a foundation for continuous process improvement and optimization, making the milling process more adaptive to varying operational conditions.

## 4.2 Control feedback

Control feedback is the final step in the entire closed-loop system and a key component of the milling DT, enabling continuous process optimization through real-time perception, learning, and control. Milling parameters change after the decision is made by the actuator, and the real-time changes in parameters such as vibration, cutting force, and tool condition occurring during the machining process are collected back into the closed loop by advanced sensing technologies such as accelerometers, acoustic emission sensors, and force sensors. This feedback loop enables the process to adapt to changing conditions dynamically, ensuring efficient and precise operation and laying the foundation for ML models to predict system behavior, identify anomalies, and recommend process adjustments.

Control feedback is critical to maintaining the stability and reliability of the whole milling system. Real-time analysis and adjustment ensure that the process responds to disturbances such as tool wear, material inconsistencies, or changes in the external environment. This dynamic adaptation prevents inefficiencies, reduces waste, and minimizes the risk of defects, directly contributing to improved product quality and system performance[132]. In addition, integrating control feedback into the sensing-learning-control loop supports active decision-making.

Control feedback plays a vital role in enabling real-time process optimization in smart milling systems. It facilitates the integration of perception, learning, and control, allowing the system to adapt to changing conditions. Control feedback improves efficiency, stability, and reliability across the production process by continuously monitoring and adjusting operations. It is a fundamental component in achieving intelligent and responsive milling.

In conclusion, Figure 5 presents a control architecture representing the interaction flow from NC codes generation to real-time control feedback. After decision-making, the translated NC codes are transmitted via standardized communication protocols or APIs to the machine controller, where they are interpreted and executed. The adjustment execution module implements these commands in real time, while integrated sensors continuously monitor the machining process and generate control feedback. This feedback enables dynamic updates to the controller, supporting adaptive and intelligent milling operations. Together, these components establish an adaptive, closed-loop control framework essential for implementing DT technologies within smart manufacturing environments.

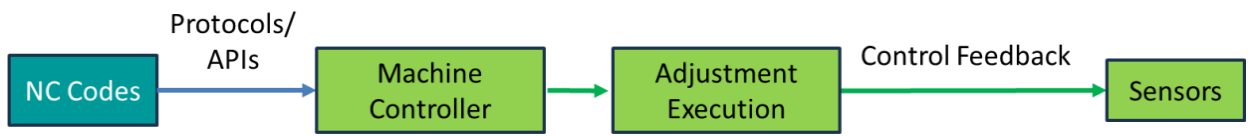

Fig. 5: Adaptive control system with sensor-based feedback

## 5. Case Study: Functional Milling DT

A meaningful review paper would have three components: analysis and synthesis of state-of-the-art, case studies to highlight key concepts or ideas, and future perspectives. Therefore, a case study was provided to highlight the concept of an AI-driven DT and its capabilities. This case study aims to demonstrate proof of concept of an intelligent DT (NN model in this case) with real-time sensing-learning-control function.

This case study showcases an AI-driven DT of a micro milling process to demonstrate how to create a functional DT. The DT aims to achieve real-time monitoring and prediction of tool-workpiece contact during micro milling by integrating real-time AE data collection, streaming, and ML predictions. The purpose of the single-modality case study is to demonstrate proof of concept of an intelligent DT, i.e., the live data stream synchronizes a DT that powers real-time control to the physical system. The case study can be extended to multi-modal manufacturing applications, which will be an important future research topic.

The DT has four key elements (Figure 6), i.e., the physical milling system - tool/workpiece interaction, the digital replica - a virtual model of the physical milling system, live data stream - sensors providing real-time input to drive the DT, and feedback control - a real-time virtual model predictive control to the physical system. The milling DT aims to achieve real-time monitoring and prediction of tool-work contact detection by integrating real-time AE data collection, streaming, and ML predictions. The key DT elements are described as follows.

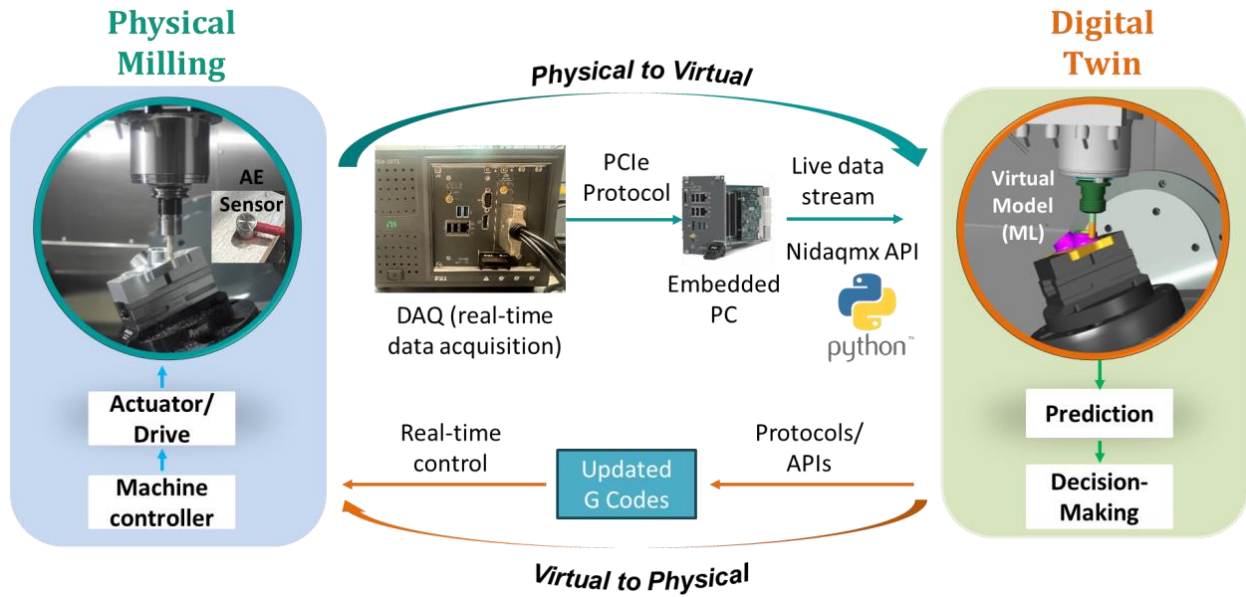

Fig. 6: Real-time AI-driven milling digital twin.

**Physical milling system:** The milling platform includes a benchtop milling machine, an AE sensor attached to the workpiece surface, a preamplifier to boost weak AE signals before digitization, and a DAQ system to collect and preprocess the real-time AE signals. A Python-based framework manages real-time data acquisition, feature extraction, and prediction using the NI-DAQmx Python API. AE signals are continuously sampled at 100 kHz and stored in a double-buffered memory to prevent data loss. The raw signals were segmented into 0.1-second windows (10,000 samples), and the peak amplitude was extracted from each window as the key feature for classification due to its simplicity and strong correlation with contact events. Each window is labeled manually based on the tool's state (contact or no contact) and used to build a structured dataset. This dataset forms the foundation for training an NN model capable of predicting real-time tool-workpiece contact detection.

**Virtual model training:** The milling process DT, an NN model, was developed to monitor the tool-work contact condition using AE peak amplitudes. A total of 1,555 labeled samples were collected and split into 80% training data (1,244 samples) and 20% test data (311 samples). The NN was designed to predict tool-workpiece contact based on AE signal peak amplitude as the input feature. The NN architecture consisted of one input neuron (peak amplitude), three hidden layers (16, 16, 8 neurons), and one output neuron for classification. All hidden layers used the ReLU activation function, while the output layer used a sigmoid activation to produce a probability of contact. The NN model was trained using the Adam optimizer with a learning rate of 0.001. Binary cross-entropy was employed as the loss function. Training was carried out for 200 epochs, with early stopping applied if validation loss did not improve for 15 consecutive epochs. The network test accuracy is 99.86%, demonstrating near-perfect classification on the testing dataset.

**Flow from physical milling to virtual model:** A real-time data streaming framework was established to transmit digital AE signals from the analog-to-digital converter for signal analysis and ML ingestion. Analog signals from the AE sensor were first amplified and digitized at a 100 kHz sampling rate, then transferred to the embedded controller via the PCI Express protocol. A Python script was developed using the NI-DAQmx API and NI-DAQmx driver to manage continuous data acquisition with double buffering to prevent data loss. The peak amplitude was extracted from the AE signal at regular intervals and fed as the primary input into the pretrained NN model. A graphical user interface was built using Python to display real-time classification results, indicating the contact status between the cutting tool and workpiece. This system architecture separates data acquisition, feature extraction, classification, and communication into synchronized processes. This structure ensures reliable performance and supports future integration with closed-loop real-time machining control.

**Feedback from virtual model to physical milling:** To enable responsive control, the system integrates real-time classification results into a feedback loop that adjusts machine behavior based on tool-work contact status. Once a contact event is detected, the updated decision is converted into modified NC codes and sent to the machine controller through communication between the software and machine layers via control APIs. The controller interprets the updated instructions and commands the machine actuator to implement the corresponding mechanical adjustments, such as modifying spindle speed or feed rate. These adjustments are reflected in the physical milling process, and the resulting changes in milling process dynamics are captured again by the AE sensor for continued monitoring. This continuous prediction, decision-making, and physical adjustment loop forms a closed-loop control strategy. It ensures the machining process adapts in real time to reduce potential machine damage risk and support intelligent automation in modern manufacturing systems.

The end-to-end real-time sensing-learning-control latency was estimated at ~10 milliseconds. A systematic measurement and comparison of latency against other approaches is beyond the scope of this work, but will be pursued in future work.

## 6. Conclusion and Outlook

### 6.1 Conclusions

This work provides a DT framework with a case study of milling processes to demonstrate the proof of concept. The DT framework allows continuous synchronization between a physical manufacturing process and its virtual model, forming a closed sensing-learning-control loop that can perceive, predict, and respond within milliseconds. Key results are summarized as follows:

(1) An intelligent DT has four essential components – A physical process, machine, or system; a virtual model (e.g., simulations, AI/ML models) of the physical entity; a live data stream to synchronize the virtual model with its physical entity in real-time; and a real-time virtual model predictive control to the physical system.

(2) Data flow from physical milling to virtual model – Data acquisition and live data streaming protocols are analyzed to update and calibrate a DT for real-time synchronization.

(3) Virtual models of physical systems – The evolution of conventional physics-based models, data-driven ML models, and physics-informed ML models is scrutinized in terms of their pros and cons.

(4) Feedback from virtual model to physical milling – A DT-powered adaptive process can be achieved through an API to translate model predictions into real-time control actions with low end-to-end latency.

This study shows that integrating real-time monitoring, live data stream, AI models, and model predictive control through the digital twin framework enables static models to become dynamically adaptive, low-latency, and autonomous models, marking a transformative approach for future smart manufacturing.

## 6.2 Future research perspectives

DTs are still at an early stage as a unanimous definition is not well defined. The DT community faces many challenges to achieve the goal of self-learning, self-adaptivity, and self-optimization autonomous manufacturing systems. Key challenges and potential research directions are outlined as follows.

- Ultra-low latency sensing and communication network – Smart manufacturing vertical-driven next generation sensing and communication network (e.g., 5G/6G) will significantly reduce latency for live data streaming and real-time control.
- Leveraging edge-cloud continuum – Future DTs will leverage edge computing for low-latency decision-making or process control and cloud computing for computation-intensive data analytics. This hybridization will enhance real-time responsiveness and scalability.
- Autonomous and self-evolving DTs – Next-generation DTs will integrate self-learning AI models to synchronize with real-world dynamics without human intervention continuously.
- Robust and resilient DTs – Future DT models should account for inherent uncertainties and unexpected disruptions in manufacturing systems.
- Data-efficient, generalizable, and interpretable DTs – Physics-informed neural networks (PINNs) may combine physical laws with small training data for high computation efficiency and model interpretability.
- Fast multiphysics simulation-based DT models – Physics-informed reduced-order modeling, integrating the accuracy of traditional numerical methods with the efficiency of the data-driven PINN approach.
- End-to-end latency – A sum of latencies from data acquisition and streaming, DT model computation, and edge-based model predictive control can be reduced at each stage of a DT ecosystem.
- Semantic interoperability and standardization – A significant challenge is ensuring semantic interoperability across diverse AI-driven DT systems with universal communication protocols (e.g., OPC UA).
- Human-DT collaboration – AI-driven twins will evolve into cognitive partners interacting naturally with humans, providing contextualized insights and predictive decision support.
- Cybersecurity and trustworthy DTs – AI-driven DTs will need trust, transparency, and explainability layers to mitigate risks from data manipulation and adversarial attacks.

## Acknowledgment

The authors would like to thank the National Science Foundation grant #2328260 for the financial support.